\documentstyle[prc,preprint,aps,epsfig]{revtex}

\newcommand{\vsig}{\mbox{\boldmath $\sigma$ \unboldmath}}
\newcommand{\veps}{\mbox{\boldmath $\epsilon$ \unboldmath}}

\begin{document}
\title{Quark model approach to the $\eta$ meson electroproduction \\
on the proton}
\author{Qiang Zhao$^{1,2}$\thanks{
Electronic address: qiang.zhao@surrey.ac.uk}, 
Bijan Saghai$^3$, and Zhenping Li$^4$}
\address{1) Department of Physics, University of Surrey, 
Guildford, Surrey GU2 7XH, United Kingdom}
\address{2) Institut de Physique Nucl\'eaire, F-91406 Orsay 
Cedex, France}
\address{3) Service de Physique Nucl\'eaire, DSM/DAPNIA, 
CEA/Saclay, \\ F-91191 Gif-sur-Yvette, France}
\address{4) Department of Physics, Peking University, 100871 Beijing, P.R. 
China}
%
%

\maketitle  
  
\begin{abstract}
A recently developed quark model approach 
to pseudoscalar meson photoproduction 
is extended to electroproduction process for the
$\eta$ meson in the kinematics  
of momentum transfer
$Q^2 \leq$ 4 (GeV/c)$^2$ and total center of mass energy $W \leq$
1.6 GeV. 
Existing data are well reproduced and the roles of the 
$S_{11}(1535)$ and $D_{13}(1520)$ resonances are closely investigated. 
In the study of the longitudinal excitation of the $S_{11}(1535)$
resonance, a reliable constraint on the $S_{11}(1535)$ properties 
is obtained 
by cleanly removing the electromagnetic transition 
from the 
$\gamma_{(v)} p \to S_{11}(1535) \to \eta p$ amplitude.
Thus, the fitted quantities can be determined 
with an uncertainty of about 15\%.
This could be the first direct constraint
on the $S_{11}(1535)$ properties in theory.

\end{abstract}
\vskip 1.cm

{PACS number(s): 12.39.-x, 13.60.-r, 13.60.Le, 14.20.Gk }

%

\newpage

\section{Introduction}

Our knowledge about the 
internal structures of baryon resonances,
which 
generally  belong to non-perturbative QCD phenomena, 
are still far from complete. Nowadays,
with the availabilities of high intensity photon and electron beams
at JLab, ESRF, MAMI, ELSA, and SPring-8, the baryon resonances
thus can be systematically investigated via meson 
photo- and electroproduction. 
This initiates various theoretical efforts, through which 
one expects that our knowledge about 
these non-perturbative phenomena can be established
on a more solid and fundamental ground.

In the late 1990's, experimental data with unprecedent accuracies were available
for the $\eta$ meson photoproduction which
allowed a close study of resonance excitations in $\gamma_{(v)} p\to \eta p$.
For example, the photoproduction data for the cross
sections and single polarization observables
obtained at MAMI~\cite{mainz}, ELSA~\cite{elsa} and 
ESRF (GRAAL)~\cite{graal,graal-coll}, had allowed to 
establish 
that the $\eta$ photoproduction reaction
mechanism was dominated by the $S_{11}(1535)$ excitation, while the
$D_{13}(1520)$ played a small but non-negligible role
~\cite{tiator95,li-eta-95,RPI96,li-saghai98,RPI98,tiator99,RPI00}. 
The measurements for polarization
observables also made it possible to single out the 
contribution of the $F_{15}(1680)$ resonance~\cite{li-saghai98}. 
However, detailed properties concerning the $\eta NN^*$ couplings
and even the $\eta NN$ coupling still could not be well-constrained
at that time, including 
the total decay width of the dominant $S_{11}(1535)$,
and its branching ratio into $\eta N$.
Such a situation somehow stalked theoretical progresses due to 
lack of constrained inputs for the modellings, although various 
QCD inspired phenomenologies, such as the quark model 
approaches~\cite{li-eta-95,li97,Pfeil,Carlson,li-saghai98}, 
and formalisms
based on the meson-nucleon degrees of freedom using isobaric  
descriptions~\cite{RPI96,RPI98,RPI00}, or
multipole analyses~\cite{tiator95,tiator99}, were proposed to
investigate the roles played by the intermediate resonances
in the $\eta$ photoproduction.

The situation was changed significantly with the electroproduction
data available from JLab~\cite{armstrong99}, which brought rich complementary 
information about the $S_{11}(1535)$ properties and the $D$-wave influences,
due to the presence of the longitudinal photon excitations.
Such an observable is not only valuable for model-selection, but also
useful for disentangling model-dependent or model-independent aspects
within a phenomenology.

In this paper, we shall study the $\eta$ meson electroproduction
in a constituent quark model with a chiral effective Lagrangian 
for the quark-meson ($\eta$) couplings, and focus our attentions on
the dominant $S_{11}(1535)$ and $D_{13}(1520)$ resonance.
The formalism is discussed in Sec.~II. In Sec.~III, we 
present detailed results produced by our model fits. 
This will allow us to proceed to a successful constraint
on the $S_{11}(1535)$ properties.
The conclusions are given in Sec.~IV.


\section{Theoretical frame}
In this work, we follow the scheme of a recently developed quark model 
approach to 
the pseudoscalar meson photoproduction~\cite{li97}, 
and extend it to the $\eta$ meson electroproduction in the region of 
the c.m. energy 
$W=1.535$ GeV. Compared to an isobaric approach, 
in this model the intermediate baryon resonances can be systematically 
included in the formalism. Starting from the NRCQM~\cite{isgur-karl}, only a limited 
number of adjustable parameters will appear in the model.

\subsection{Formalism}

For the meson interaction vertex, a QCD inspired 
effective Lagrangian~\cite{MANOHAR} is introduced to account for the 
quark-meson interaction:
\begin{equation}\label{lagrangian}
L_{eff}=\sum_{j}\frac{1}{f_\eta}\overline{\psi}_j\gamma_\mu^j
\gamma_5^j\psi_j\partial^\mu\phi_\eta \ ,
\end{equation}
where $\psi_j$ ($\overline{\psi}_j$) is the $j$th quark (anti-quark) field
in the nucleon, and $\phi_\eta$ represents the $\eta$ meson field. 
At quark level, the transition matrix element ${\cal M}_{fi}$
can be expressed as the sum over the {\it t}-, {\it s}- and {\it u}-channel
transitions, i.e. ${\cal M}_{fi}={\cal M}_{fi}^t
+{\cal M}_{fi}^s+{\cal M}_{fi}^u$. 
Since ${\cal M}_{fi}^t$ is proportional to the
charge of the outgoing meson, it vanishes in the neutral meson production. 
In the {\it s}- and {\it u}-channel, 
contributions from a complete set of intermediate 
baryon resonances are included in the quark model 
SU(6)$\otimes$O(3) symmetry limit. It can be seen easily 
that due to the isospin conservation, isospin 3/2 states cannot 
contribute in this channel. Nevertheless, the Moorhouse selection
rule eliminates the states of representation $[{\bf 70}, \ ^{\bf 4} {\bf 8}]$
from contributing to the proton target reaction~\cite{moorhouse}.  
Thus, eight isospin-1/2 resonances corresponding to 
the harmonic oscillator quantum numbers $n$=1 and~2,
and the nucleon pole terms, are to be included explicitly. 
States with $n\ge 2$ are treated as degenerate in the quantum number $n$.

Adopting Lorentz gauge $k_\mu A^\mu=0$, 
the longitudinal virtual photon polarization vector is defined as 
\begin{equation}
\varepsilon_\mu^L =\frac{1}{\sqrt{Q^2}} \left( |{\bf k}|, \ 0, \ 0, \ 
\omega_\gamma \right ) \ ,
\end{equation}
where $|{\bf k}|$ and $\omega_\gamma$ are
 the momentum and energy of the virtual photon in the meson-nucleon
c.m. frame, respectively.  
Gauge invariance requires that the longitudinal transition 
operator is proportional to $\sqrt{Q^2}/|{\bf k}|$. Therefore, the 
longitudinal transition vanishes in the real photon limit. 
The longitudinal electromagnetic interaction is defined as
\begin{equation}
H_{em}^L=\varepsilon_0^L J_0 -\varepsilon_3^L J_3 \ , 
\end{equation}
where the electromagnetic current components
$J_0$ and $J_3$ for a three-quark system can be expressed as
\begin{equation}
J_0=\sqrt{4\pi}\frac{1}{\sqrt{2\omega_\gamma}} \sum_{j=1}^3 
e_j e^{i{\bf k}\cdot {\bf r}_j} \ , 
\end{equation}
and
\begin{equation}
J_3=\sqrt{4\pi}\frac{1}{\sqrt{2\omega_\gamma}} \sum_{j=1}^3 e_j \alpha_j
e^{i{\bf k}\cdot {\bf r}_j} \ ,
\end{equation}
with $e_j$ denotes the charge operator of the $j$th quark, and $\alpha_j$, 
the Dirac matrix. Given the Hamiltonian of the three-quark 
system described by only two-body interactions~\cite{McClary83}, 
the current conservation gives
\begin{equation}
\langle N_f| J_3 | N_i\rangle = \frac{E_f-E_i}{k_3} 
\langle N_f| J_0 | N_i\rangle \ ,
\end{equation}
where $k_3=|{\bf k}|$ and the energy conservation gives $E_f-E_i=\omega_\gamma$.  
The longitudinal interaction can then be written as
\begin{equation}
H_{em}^L=\left[ \varepsilon_0^L-\varepsilon_3^L
\frac{\omega_\gamma}{k_3}\right ] J_0 \ .
\end{equation}

Substituting the $\varepsilon_\mu^L$ into the above equation, one obtains
the gauge-invariant relation: 
\begin{equation}
\langle N_f | H_{em}^L | N_i \rangle = \sum_{j=1}^3 
\frac{\sqrt{Q^2}}{|{\bf k}|}\langle N_f | J_0 | N_i\rangle \ .
\end{equation}

We adopt the following nonrelativistic expansion of $J_0$ for the 
longitudinal transition~\cite{close-li-92}:
\begin{eqnarray}
\label{longi-operator}
J_0^{NR} & = & \sqrt{4\pi} \frac{1}{\sqrt{2\omega_\gamma}} \{
\sum_{j} \left[ e_j + \frac{ie_j}{4m_j^2}
{\bf k}\cdot (\vsig_j\times {\bf p}_j) \right] 
e^{i{\bf k}\cdot {\bf r}_j}\nonumber\\
& & -\sum_{j<l} \frac{i}{4M_T}(\frac{\vsig_j}{m_j}-\frac{\vsig_l}{m_l})
\cdot \left[ e_j {\bf k}\times {\bf p}_l e^{i{\bf k}\cdot {\bf r}_j}
-e_l {\bf k}\times {\bf p}_j e^{i{\bf k}\cdot {\bf r}_l} \right] \} \ ,
\end{eqnarray}
where the first term describes the c.m. motion of the three-quark system
and is commonly used in the longitudinal transitions. 
The second and third terms are spin-orbit and non-additive terms, 
which are believed to be the leading order relativistic corrections 
to the transition operators~\cite{li-close-90}. 
We shall neglect the spin-orbit and non-additive terms in this work.
It can be seen later that this approximation 
will result in an over-estimation of the 
$S_{11}(1535)$ longitudinal excitation cross section. 
Empirically, a coefficient for the $S_{11}$ longitudinal 
transition amplitude will be introduced. 

The transverse transition amplitude for the photoproduction 
have been derived in Ref.~\cite{li97}, in which the transition amplitudes 
for the $S_{11}(1535)$ and $D_{13}(1520)$ are 
\begin{equation}
\label{S11-transverse}
{\cal M}_{S_{11}}=
\frac{2M_{S_{11}} e^{-\frac{{\bf k}^2+{\bf q}^2}{6\alpha^2}} }
{(s-M_{S_{11}}^2+i M_{S_{11}}\Gamma_{S_{11}})}
\frac{1}{6} \left( \frac{\omega_\eta}{\mu_q}-(\frac{\omega_\eta}{E_f+M_N}+1)
\frac{2{\bf q}^2}{3\alpha^2} \right) (\omega_\gamma+\frac{{\bf k}^2}{2m_q})
\vsig\cdot\veps_\gamma \ ,
\end{equation}
and
\begin{eqnarray}
\label{D13-transverse}
{\cal M}_{D_{13}}&=&\frac{2M_{D_{13}} 
e^{-\frac{{\bf k}^2+{\bf q}^2}{6\alpha^2}}}
{(s-M_{D_{13}}^2+i M_{D_{13}}\Gamma_{D_{13}})}
\{ (\frac{\omega_\eta}{E_f+M_N}+1)\frac{{\bf q}^2}{9\alpha^2}
(\omega_\gamma+\frac{{\bf k}^2}{2m_q}) \vsig\cdot\veps_\gamma \nonumber\\
&& -\frac{i}{2m_q}(\frac{\omega_\eta}{E_f+M_N}+1) 
\frac{{\bf k}\cdot{\bf q}}{3\alpha^2} \vsig\cdot {\bf q} 
\vsig\cdot ({\bf k}\times\veps_\gamma) \nonumber\\
&& -(\frac{\omega_\eta}{E_f+M_N}+1)\frac{\omega_\gamma}{3\alpha^2}
\vsig\cdot {\bf q}\veps_\gamma\cdot{\bf q} \} \ ,
\end{eqnarray}
where $\mu_q$ is the reduced mass of two quarks and equals $m_q/2$
in the $\eta$ production for $m_u=m_d=m_q$. 
The overall coupling $\alpha_\eta$ is related to the $\eta$ meson 
radiative decay constant
$f_\eta$ 
by assuming the validity of the Goldberger-Treiman
relation~\cite{goldberger-treiman} for the $\eta NN$ couplings: 
\begin{equation}
g_{\eta NN}=\frac{g_A M_N}{f_\eta} \ ,
\end{equation}
with $\alpha_\eta\equiv g_{\eta NN}^2/4\pi$, where $g_A=1$ is given
by the NRCQM in the SU(6)$\otimes$O(3) symmetry limit.

Similarly, the amplitudes for other resonances in the quark model can 
be derived. In this paper, we  
concentrate on the kinematics 
from threshold to $W\sim 1.54$ GeV, where 
the available experimental data 
allow a close study of the $S_{11}(1535)$ and $D_{13}(1520)$. 
One of the advantages is that one can neglect the SU(6)$\otimes$O(3)
symmetry violations for other excited states. 
Therefore, we could only introduce an 
SU(6)$\otimes$O(3) beaking coefficient for 
the $S_{11}(1535)$ and $D_{13}(1520)$.
In another word, except for the $S_{11}(1535)$ and $D_{13}(1520)$,
the relative strengths among other resonances will be constrained 
by the quark model. In this way, we avoid to introduce too many 
parameters in the model. 
This treatment is different from that in Refs.~\cite{li-saghai98}
and \cite{saghai-li-01}. 
Another feature in this calculation is that the quark model parameters
are also taken into account. The harmonic oscillator strength $\alpha$
is treated as a free parameter, of which 
a value within the range of quark model validity could be an essential
test of the 
self-consistence of this model. 
Moreover, the $S_{11}(1535)$ total width $\Gamma_{S_{11}}$ 
and its partial decay
branching ratio $b_\eta$ into $\eta N$ will
be treated as free parameters as well.

\subsection{Kinematics and observables}
The kinematics of the meson exclusive electroproduction has been discussed 
in the literature~\cite{garcilazo93,lee90}. 
Here, we directly relate the kinematics to our 
convention of the transition amplitude. 
Generally, the cross section of the meson electroproduction  
can be written as follows:
\begin{equation}
\frac{d\sigma}{dE^\prime_e d\Omega_2 d\Omega^*} = 
\Gamma_v \frac{d\sigma}{d\Omega^*} \ ,
\end{equation}
where, $d\sigma/d\Omega^*$ represents the cross section for the 
virtual-photon-nucleon scattering in the meson-nucleon c.m. frame, 
while the contribution from the lepton current can be factorized into 
a factor $\Gamma_v$ 
which is known as the virtual photon flux 
\begin{equation}
\Gamma_v=\frac{\alpha_e}{2\pi^2}\frac{E^\prime_e}{E_e}\frac{K_E}{Q^2}
\frac{1}{1-\varepsilon} \ ,
\end{equation}
where $\alpha_e$ 
is the electromagnetic fine structure constant, 
$E_e$ and $E^\prime_e$ are the energies of the initial and scattered electrons,
respectively, in the lab system, and $K_E=(s-M^2)/2M$ is the equivalent 
energy of the virtual photon as a real photon in the lab system. 

The virtual photon polarization parameter, $\varepsilon$, is defined as
\begin{equation}
\varepsilon=(1+2\tan ^2(\frac{\theta_e}{2})|{\bf k_0}|^2/Q^2)^{-1} \ .
\end{equation}
where ${\bf k_0}$ is the virtual photon momentum in the lab system.

The cross section for the meson production by the virtual photon 
in the meson-nucleon c.m. frame can be expressed as 
\begin{equation}
\label{diff-xsect}
\frac{d\sigma}{d\Omega^*} = \frac{d\sigma_T}{d\Omega^*}
+ \varepsilon\frac{d\sigma_L}{d\Omega^*}
-\varepsilon \frac{d\sigma_{TT}}{d\Omega^*}\cos{2\phi_\eta^*}
-\sqrt{\varepsilon(1+\varepsilon)}\frac{d\sigma_{TL}}{d\Omega^*}
\cos{\phi_\eta^*} \ , 
\end{equation}
where $\phi_\eta^*$ is the azimuthal angle between the $\eta N$
scattering plane and the $(e, e^\prime)$ scattering plane, 
and 
\begin{eqnarray}
\frac{d\sigma_T}{d\Omega^*}&=&\xi \ 
\frac{2 |{\bf k}|^2}{Q^2} {\cal H}^{00} \ , \\
\frac{d\sigma_L}{d\Omega^*}&=&\xi \ ({\cal H}^{11}-{\cal H}^{-1-1}) \ , \\
\frac{d\sigma_{TT}}{d\Omega^*}&=&\xi \ ({\cal H}^{1-1}+{\cal H}^{-1 1}) \ , \\
\frac{d\sigma_{TL}}{d\Omega^*}&=&\xi \ \sqrt{\frac{|{\bf k}|^2}{Q^2}}
({\cal H}^{01}+{\cal H}^{10}-{\cal H}^{0-1}-{\cal H}^{-10}) \ .
\end{eqnarray}
In the above equations, $\xi$ represents the phase space factor 
in the virtual-photon-nucleon excitations, and has the following 
expression:
\begin{equation}
\xi=\frac{\alpha_e}{16\pi}\frac{M_N}{W}\frac{|{\bf q}|}{K_E} \ ,
\end{equation}
where $|{\bf q}|=\left[ (s-M_N^2-m_\eta^2)^2
-4M_N^2m_\eta^2\right ]^\frac12/(2W)$ 
is the meson momentum in the meson-nucleon c.m. frame.
The amplitude,
${\cal H}^{\lambda\lambda^\prime}$, is defined as
\begin{equation}
{\cal H}^{\lambda\lambda^\prime}=
\sum_{\lambda_1 \lambda_2}{\cal A}^\lambda_{\lambda_1 \lambda_2}
{{\cal A}^{\lambda^\prime}_{\lambda_1 \lambda_2}}^\dag \ ,
\end{equation}
where ${\cal A}^\lambda_{\lambda_1 \lambda_2}$ are the transition 
amplitudes for the meson production via the virtual photon, 
$\lambda=0, \pm 1$ denotes the spin polarizations 
of the incoming virtual photon and, $\lambda_1=\pm \frac 12$ and 
$\lambda_2=\pm \frac 12$ for 
the initial and final state nucleons, respectively. 
In the pseudoscalar meson electroproduction, there are 
two independent amplitudes in the longitudinal 
photon transitions, and four independent 
 amplitudes in the transverse photon transitions.
The latter ones can be related 
to the traditional CGLN amplitudes 
or the helicity amplitudes~\cite{fasano92}.

When the calculations are extended 
to the electroproduction, it becomes highly relativistic, and 
a Lorentz boost factor must be introduced into the spatial integrals. 
We follow the prescription of Foster and Hughes~\cite{fostor82} 
to boost momentum for the spatial integrals 
in an equal velocity frame (EVF), which 
is very close to the Breit frame. 
In the EVF, the Lorentz boost factor for the virtual photon 
interaction is defined as 
\begin{equation}
\label{EVF-photon}
\gamma_k=\left( 1+\frac{k^2}{(W+M_N)^2} \right)^{\frac 12} \ , 
\end{equation}
where the relation of the momentum $k$ with the $Q^2$ in the EVF is 
\begin{equation}
\label{k2}
k^2(EVF)=\frac{(W^2-M_N^2)^2}{4W M_N} + \frac{Q^2(W+M_N)^2}{4W M_N} \ .
\end{equation}
With the boost factor, the spatial integral concerning the photon excitation 
is boosted as
\begin{equation}
R(k)\to \frac{1}{\gamma_k^2}R(\frac{k}{\gamma_k}).
\end{equation}

The same prescription can be adopted for the resonances decaying into the
final state $\eta$ meson and proton:
\begin{equation}
\gamma_q=\left( 1+\frac{q^2}{(W+M_N)^2} \right)^{\frac 12} \ , 
\end{equation}
where 
\begin{equation}
q^2(EVF)=\frac{(W^2-M_N^2)^2}{4W M_N} - \frac{m_\eta^2(W+M_N)^2}{4W M_N} \ .
\end{equation}

Therefore, the spatial integrals in the transition matrix elements from 
the initial states to the final states can be expressed as 
\begin{equation}
\label{EVF-boost}
R(k,q)\to \frac{1}{\gamma_k^2}\frac{1}{\gamma_q^2}
R(\frac{k}{\gamma_k},\frac{q}{\gamma_q}).
\end{equation}

Compared to the previous study of the $\eta$ meson 
photoproduction~\cite{li-saghai98}, in which the Lorentz boost factors,
$\gamma_k=E_i/M_N$ and $\gamma_q=E_f/M_f$, in the c.m. frame of the 
meson-nucleon system were adopted,
the Lorentz boost factors in the EVF turn out to be more successful in 
reproducing the data. 
The failure of the Lorentz boost factor $\gamma_k$ in the meson-nucleon c.m.
frame 
can be seen more clearly by rewriting it in the electroproduction:
\begin{equation}
\label{breit-photon}
\frac{1}{\gamma_k^2}=\left ( \frac{M_N}{E_i} \right )^2 
=\frac{4s M_N^2}{(s+M_N^2)^2}\frac{1}{(1+Q^2/(s+M_N^2))^2} \ .
\end{equation}
Since $4s M_N^2/(s+M_N^2)^2 \le 1$, the rest part plays 
a role as a dipole in the nucleon's electromagnetic form factor. 
Note that, the c.m. energy $W\equiv \sqrt{s}$ is fixed at $M_{S_{11}}$, 
therefore, $(s+M_N^2)$ is a constant.
In contrast to the Lorentz boost factor in the EVF, 
an {\it ad hoc} dipole, $1/(1+Q^2/0.71)^2$,
was widely used for the nucleon's electromagnetic form factor 
in the literature~\cite{ravndal71}.
Such a $Q^2$-dependence results in a fast drop-down behavior,
which underestimates the data significantly at high $Q^2$ regions.

\begin{figure}
\begin{center}
\epsfig{file=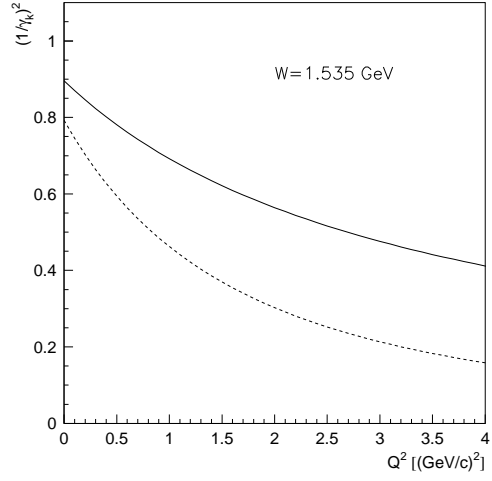,height=8.0cm,width=8.0cm}
\end{center}
\caption{ The $Q^2$-dependence of the Lorentz boost factor
in the EVF (solid) and meson-nucleon c.m. frame (dashed), respectively.
 }
\protect\label{fig:(1)}
\end{figure}

To show the difference between the two Lorentz boost schemes, 
we present the $Q^2$-dependence of the $1/\gamma_k^2$ factor
calculated at $W=1.535$ GeV
by Eqs.~\ref{EVF-photon} (solid curve) and ~\ref{breit-photon} 
(dashed curve) in Fig.~\ref{fig:(1)}. It shows that 
in the EVF, electromagnetic transition is boosted 
rather moderately than in the meson-nucleon c.m. frame. 
The difference between these two frames reveals non-covariance  
of this approach. As the boost goes large, a more realistic prescription
is needed.

For the meson interaction vertex, the boost factor at $W=1.535$ GeV
in both frames is a constant, and possesses very close values, 
i.e., $1/\gamma_q^2=0.99$ in the EVF and 0.96 in the meson-nucleon
c.m. frame. Such a feature is due to the quite heavy mass 
of the $\eta$ meson 
and its relatively small on-mass-shell momentum.

\section{Results and discussions}

In this Section, we present numerical results for
various photo- and electroproduction observables.
Parameters are fitted by experimental data.
Meanwhile, an analytical relation is deduced 
in the study of the $S_{11}(1535)$ excitations.

\subsection{Database and adjustable parameters}

The newly published data~\cite{armstrong99}
cover the kinematics   
$1.490 \leq W \leq 1.615$~GeV at $Q^2=$2.4 and 3.6 (GeV/c)$^2$, 
have greatly improved 
the experimental status of the $\eta$ meson photo- and electroproduction.
Along with three other datum sets:  
old electroproduction data at W=1.535 GeV with 
$Q^2=$0.20, 0.28 and 0.40 (GeV/c)$^2$~\cite{beck74}, 
and recent photoproduction
cross section data from Mainz~\cite{mainz} and GRAAL~\cite{graal-coll}
from $\eta N$ threshold to $E_{\gamma}^{lab} \leq 0.9$
GeV (corresponding to $W=1.6$ GeV), we obtain a large database 
covering the complete kinematics of the $S_{11}(1535)$ 
and $D_{13}(1520)$ excitations. This opportunity will 
allow us to concentrate on these two resonances, 
whose couplings to the $\eta N$ channel have not been well known.
Note that, the Mainz data cover the energy range from threshold to 
$E_{\gamma}^{lab}$=0.789 GeV, while the GRAAL data go to higher
energies: $0.714 <E_{\gamma}^{lab} \leq 1.1$ GeV. For
GRAAL data we use a subset restricted to $E_{\gamma}^{lab} \leq 0.9$
GeV.

The complete database contains 677 differential cross-sections and 
has been used to
fix 7 adjustable parameters in this approach.
The results, obtained using the CERN MINUIT minimization code,
are given in Table~\ref{tab:(1)}. 
Below, we summarize the main features of each parameters
based on the numerical investigations. 

The $\eta NN$ coupling constant 
($\alpha_{\eta} \equiv g_{\eta NN}^2/4\pi$)
is not a well-determined quantity in the literatures. 
Here, $\alpha_{\eta}$ is an overall factor and the numerical 
fits give
$\alpha_{\eta}=0.10\pm 0.01$.
Recent studies~\cite{li-saghai98,cc,zhu} suggest that 
$g_{\eta NN}^2/4 \pi$ is much smaller than unit.

The commonly used value for the second free parameter, namely, the 
harmonic oscillator 
strength $\alpha$ varies between 330 and 430 MeV in the literatures. 
In the numerical fitting, 
a rather moderate value, $\alpha=384.5$ MeV, is extracted.  
With the quark model spatial wavefunctions boosted in the EVF, 
Such a value avoids the introduction of an 
{\it ad hoc}  dipole or monopole in the form factor~\footnote{
In Ref.~\cite{fostor82}, 
the authors adopted a larger value for the harmonic oscillator 
strength, $\alpha=458.2$ MeV. 
Meanwhile, a monopole form factor, $F_q(Q^2)=1/(1+Q^2/0.8)$, 
had to be adopted to  
produce the dying-out behavior of the resonance helicity amplitudes
in the $Q^2$-dependence. Also, it was shown 
that a smaller value $\alpha=331.7$ MeV with $F_q(Q^2)=1$ could 
nicely reproduce the helicity amplitudes for the $S_{11}(1535)$.
See the dot-dashed curve in Fig. 1c of Ref.~\cite{fostor82}.}.
Then, the exponential factor from the spatial 
integrals, $e^{-({\bf k}^2+{\bf q}^2)/6\alpha^2}$, will play a role 
like a form factor after being boosted.

The $S_{11}(1535)$'s total width $\Gamma_{S_{11}}$ and its 
$\eta N$ decay branching ratio $b_\eta$ have been very important in the 
study of its nature. However, both experimental values
and theoretical predictions for these two 
quantities have been very contradictory
in the literatures~\cite{mainz,armstrong99,manley92,green97}. 
The $\eta$ meson photo- and electroproduction at the $S_{11}$ 
mass region show great sensitivities to these two quantities. 
Much stronger constraints from the electroproduction data 
are found in this calculation.  
Treating the two quantities as free parameters, we extract
$\Gamma_{S_{11}}=143.3\pm 0.2$ MeV and $b_\eta(S_{11})=0.55\pm 0.01$.
These results are very close to the most recent 
extraction from the JLab data~\cite{armstrong99}. 
For other resonances, their total widths and partial decay branching ratios
are found not to be sensitive in this kinematical region.
Therefore, we use the total widths
from the Particle Data Group (PDG)~\cite{PDG98} in the calculation. 
The partial decay channels for those resonances are restricted to the 
$\pi N$ and $\eta N$, with branching ratios fixed at 
$b_\pi$=90\% and $b_\eta$=10\%. 
As a sensitivity test, we also use $b_\pi$=99\% and $b_\eta$=1\% and
found that the numerical results
do not show significant dependence on these branching ratios due to the
dominance of the $S_{11}(1535)$.

The fifth and sixth parameter, $C_{S_{11}}$ and $C_{D_{13}}$,
are strength coefficients introduced for 
the $S_{11}(1535)$ and $D_{13}(1520)$ amplitudes to take care of 
the breaking of the 
SU(6)$\otimes$O(3) symmetry. 
As discussed in Sec. II,
eight resonances with harmonic oscillator shell $n\le 2$
are to be explicitly included, i.e., 
$S_{11}(1535)$, $D_{13}(1520)$, $P_{13}(1720)$,
$F_{15}(1680)$, $P_{11}(1440)$, $P_{11}(1710)$, $P_{13}(1900)$,
and $F_{15}(2000)$.
In the SU(6)$\otimes$O(3) symmetry limit, 
the relative strengths of
all these resonances and 
higher mass terms are fixed with the same
quark-meson coupling denoted by $\alpha_\eta$,
apart from the quark model couplings. 
The configuration mixing among excited states, which breaks the 
SU(6)$\otimes$O(3) symmetry, in principle
should be taken into account. However, since the complexity
arising from such a scheme somehow cannot be easily controlled, one can
take an empirical strategy to
introduce the strength coefficient $C_{N^*}$ 
for each resonance~\cite{li-saghai98,saghai-li-01}. These coefficients
can be determined by numerical fits, and their
deviations from unit will reflect 
the breaking of the SU(6)$\otimes$O(3) symmetry.
In this study, the selected kinematical region allows us to
focus on the two resonances $S_{11}(1535)$ and $D_{13}(1520)$.
Thus, only $C_{S_{11}}$ and 
$C_{D_{13}}$ are introduced and treated as parameters.
It should be noted that, coefficients $C_{S_{11}}$ and 
$C_{D_{13}}$ were found
close to unit in 
a recent photoproduction study~\cite{saghai-li-01}. 

Finally, we include an additional coefficient $C_{S_{11}}^L$ for the 
the longitudinal 
excitation amplitude of the $S_{11}$ in the 
electroproduction process.
This quantity will be discussed in the following Section with respect to the 
experimental indications~\cite{breuker78,Br-84}.


\subsection{Analysis of observables}

With those parameters (Table~\ref{tab:(1)}) extracted in the data fits,
we shall proceed to the  
detailed analyses of the model-fitting results.
Predictions thus can be made for various observables.
Some dynamical features 
arising from the electroproduction processes 
will be investigated. The $S_{11}(1535)$ properties 
can be reasonably constrained.


\subsubsection{Cross-sections}

\begin{figure}[htb]
\begin{center}
\epsfig{file=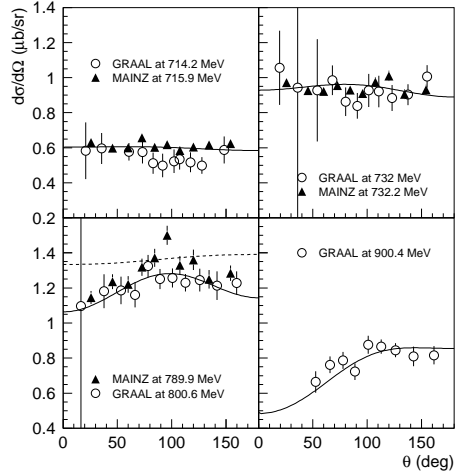,height=8.0cm,width=8.0cm}
\end{center}
\caption{
Differential cross sections in the real photon limit. Solid curves 
denote the fitting results, while the dashed curve denotes
the result in the absence of the $D_{13}(1520)$. 
Data are from Mainz~\protect\cite{mainz} (triangles) 
and GRAAL~\protect\cite{graal-coll} (circles). }
\protect\label{fig:(2)}
\end{figure}

In Fig.~\ref{fig:(2)}, the fitting 
results for the photoproduction process are shown
for energies from threshold up to $E_\gamma^{lab}=900.4$ MeV.
The full curves come from the model outlined above and reproduce
well enough the data. 
The dashed curve in Fig.~\ref{fig:(2)} at $E_\gamma=790$ MeV
denotes the result in the absence of
the $D_{13}(1520)$ contribution at the $S_{11}(1535)$ energy. 
Comparison between the full and dashed curve shows
that the $D_{13}(1520)$ plays a small but non-negligible role
in $\gamma p\to \eta p$. It accounts for the main non-$S$-wave
behavior in the angular distributions. Meanwhile, the $S_{11}$
dominance is displayed by 
the (almost) isotropic behavior of the dashed curve.

\begin{figure}
\begin{center}
\epsfig{file=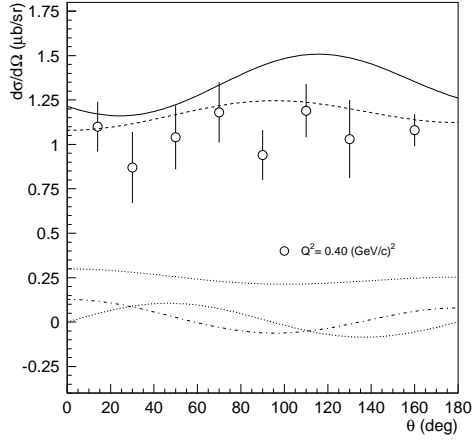,height=8.0cm,width=8.0cm}
\end{center}
\caption{ Differential cross section
 at $Q^2=0.4$ (GeV/c)$^2$ with $\phi_\eta^*=0^\circ$. 
The solid curve comes from the global fits. 
Different components
of the cross section are also presented:
the dashed curve for $d\sigma_T/d\Omega^*$; 
the dotted for $d\sigma_L/d\Omega^*$;
the dot-dashed for $d\sigma_{TT}/d\Omega^*$; 
and the heavy-dotted for
$d\sigma_{TL}/d\Omega^*$. Data are from Ref.~\protect\cite{beck74}. }
\protect\label{fig:(3)}
\end{figure}

In Fig.~\ref{fig:(3)}, the differential cross sections for the 
electroproduction
process at low-$Q^2$ are presented (full curve) and compared with the old
data at $Q^2=$0.4 (GeV/c)$^2$ with $\varepsilon=0.79$
and $\phi_\eta^*=0^\circ$~\cite{beck74}. 
Components of the cross section (Eq.~\ref{diff-xsect}) 
are also presented:
transverse (dashed curve), longitudinal (dotted curve),  
transverse-transverse (TT) interfering (dash-dotted), and 
transverse-longitudinal (TL) interfering component (heavy dotted curve).
Since the longitudinal contribution is quite small, 
it is the TT interference that
accounts for the structure at small angles as shown by the solid curve.
Meanwhile, 
it shows that the data prefer even smaller longitudinal cross sections
at this region, although 
the fits are in agreement with the data reasonably. 

\begin{figure}
\begin{center}
\epsfig{file=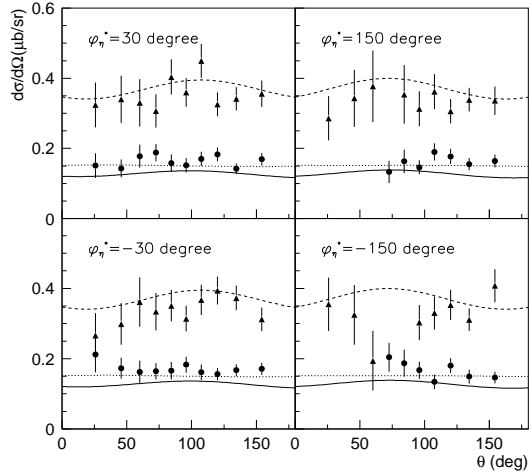,height=8.0cm,width=8.0cm}
\end{center}
\caption{ Fitting results for the differential cross sections 
at $Q^2=2.4$ (dashed)
and 3.6 (GeV/c)$^2$ (solid) for different $\phi_\eta^*$. 
The dotted curves 
are calculations without $D_{13}(1520)$ contributions at $Q^2=3.6$ (GeV/c)$^2$.
Data are from Ref.~\protect\cite{armstrong99}.
The triangles are data at $Q^2=3.6$ (GeV/c)$^2$, while
the full dottes are data at $Q^2=2.4$ (GeV/c)$^2$. }
\protect\label{fig:(4)}
\end{figure}

The recent precise measurements~\cite{armstrong99} of the $\eta$ meson 
electroproduction at $Q^2=$2.4 and 3.6 (GeV/c)$^2$ are successfully  
fitted by this approach. 
In Fig.~\ref{fig:(4)}, the angular 
distributions at $W=1.54$ GeV with different azimuthal angles $\phi_\eta^*$
for both momentum transfers are presented. 
The polarization parameters are  
$\varepsilon=0.51$ for $Q^2=2.4$ (GeV/c)$^2$ and 
0.46 for $Q^2=3.6$ (GeV/c)$^2$. 
The solid curves represent the results 
for $Q^2=3.6$ (GeV/c)$^2$ and the dashed curves for $Q^2=2.4$ (GeV/c)$^2$.
The results in the absence of 
the $D_{13}(1520)$ at $Q^2=3.6$ (GeV/c)$^2$ are shown by 
the dotted curves. 
It confirms that the non-$S$-wave 
feature of the data is due to the $D_{13}(1520)$ 
interference. Meanwhile, other resonances  
and the nucleon pole terms have only negligible influences.
At $Q^2=3.6$ (GeV/c)$^2$, the fits start to underestimate 
the experimental dottes. It might suggest that 
the NRCQM form factor for the $D_{13}(1520)$ becomes inappropriate with
the increasing $Q^2$. 
In Ref.~\cite{stoler93}, the data showed that the $D_{13}$ 
form factor had more steeper slope with the increasing $Q^2$,
which led to a negligible $D_{13}$ contribution above $Q^2=3.0$ (GeV/c)$^2$. 
However, in Fig.~\ref{fig:(4)}, 
 the $D_{13}$ still makes sense at $Q^2=3.6$ (GeV/c)$^2$, 
which means it does not decrease 
as fast as the data would require. 

\begin{figure}
\begin{center}
\epsfig{file=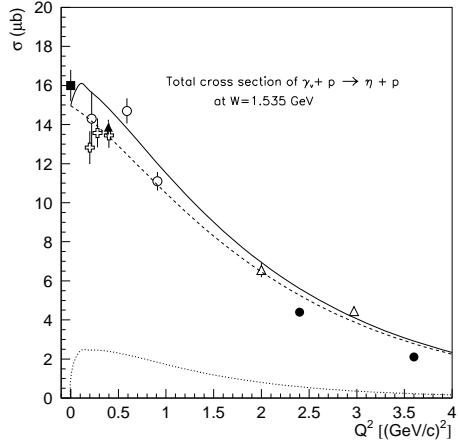,height=8.0cm,width=8.0cm}
\end{center}
\caption{
The $Q^2$-dependence of the total cross sections (solid curve). 
The total transverse and longitudinal cross sections are also shown by the 
dashed and dotted curves, respectively. 
Data come from Refs.~\protect\cite{mainz,beck74,Br-78,armstrong99,Ad-75,Br-84}. }
\protect\label{fig:(5)}
\end{figure}

With the parameters fixed,
in Fig.~\ref{fig:(5)}, the $Q^2$-dependence of the transverse and longitudinal
components of the total cross section are shown by the dashed and dotted
curves, respectively. 
We plot the total cross section
$\sigma_{tot}=\sigma_T + \varepsilon \sigma_L$ (full curve), 
at the $W= 1.535$ GeV, with $\varepsilon$=0.6. The value of the latter quantity
corresponds to the JLab experiment kinematics~\cite{armstrong99}.
The data are correctly reproduced.
Although the values of $\varepsilon$, 
determined by the experimental kinematics, varies from one datum set
to another, the chosen value for $\varepsilon$ is not crucial in the 
illustration of the calculation. This is due to the fact that 
$\sigma_L$ comes out much smaller than $\sigma_T$.

An interesting quantity to be investigated  
is the ratio $R(Q^2)=\sigma_L$/$\sigma_T$. 
Its $Q^2$-dependence 
was found to be quite smooth in both
 experiment~\cite{breuker78,Br-84,Br-78} 
and theory~\cite{ravndal71}.

As a preparation, the two well known relations 
for an excited resonance in $\gamma_{(v)} N\to\eta N$
are presented:
\begin{equation}
\label{T}
\sigma_T= 
\frac{M_N}{M_R}\frac{b_\eta}{\Gamma_R}  
2 \{|A_{\frac 12}|^2 + |A_{\frac 32}|^2 \}\ ,
\end{equation}
and 
\begin{equation}
\label{L}
\sigma_L=\frac{Q^2}{|{\bf k}|^2}
\frac{M_N}{M_R}\frac{b_\eta}{\Gamma_R}  
4 |S_{\frac 12}|^2\ ,
\end{equation}
where, $\Gamma_R$ is the total width of the resonance and 
$b_\eta\equiv \Gamma_\eta/\Gamma_R$ is the branching ratio of the
resonance decaying into $\eta N$. 

In the calculations, we find that the 
inclusive (all contributions) and exclusive (only $S_{11}$) cross sections 
produce the same ratios because 
the $S_{11}$ dominates in both $\sigma_L$ and $\sigma_T$. Then, 
the ratio can be expressed as 
\begin{equation}
R_{th}(Q^2)\equiv \frac{\sigma_L}{\sigma_T}\approx
\frac{2 Q^2}{ |{\bf k}|^2} \frac{|S_{\frac 12}|^2}{|A_{\frac 12}|^2} \ .
\end{equation}
Note that, the amplitude $A_{\frac 32}$ 
vanishes in the $S_{11}$ excitations.

\begin{figure}
\begin{center}
\epsfig{file=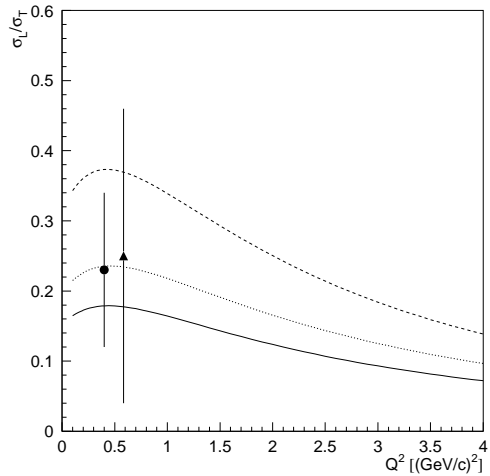,height=8.0cm,width=8.0cm}
\end{center}
\caption{
The $Q^2$-dependence of the ratio of $\sigma_L/\sigma_T$. 
The solid and dotted curves correspond to  
$C_{S_{11}}^L=0.65$ and 0.70, respectively, while
the dashed curve is for $C_{S_{11}}^L=1$. 
The full circle is from Ref.~\protect\cite{breuker78} and the triangle
from Ref.~\protect\cite{Br-84}. }
\protect\label{fig:(6)}
\end{figure}

However, in both inclusive and exclusive calculations, the present 
NRCQM study slightly overestimates  the experimental result~\cite{breuker78}. 
Empirically, we define a parameter $C_{S_{11}}^L$
for the $S_{11}$ longitudinal amplitude such that
\begin{equation}
R_{exp}(Q^2)\equiv 
(C_{S_{11}}^L)^2 R_{th}(Q^2)\ .
\end{equation}
Given $R_{exp}=0.23\pm 0.11$ at $Q^2=0.4$ (GeV/c)$^2$ 
from Ref.~\cite{breuker78},
we find 
$C_{S_{11}}^L=(R_{exp}/R_{th})^{\frac 12}=0.70$. 
This result is in good agreement with the fitted value $C_{S_{11}}^L=0.65$
in Table~\ref{tab:(1)}, and explains the introduction of this parameter 
for the $S_{11}$. 
The successful extraction of $C_{S_{11}}^L$ in the fits
 suggests that
 the leading order relativistic 
correction to the $S_{\frac 12}$ could be absorbed into the parameter 
$C_{S_{11}}^L$ and such a correction might not be sensitive to a wide $Q^2$ 
range. The latter point can be explained by the 
flat $Q^2$ dependence of the ratio.
In Fig.~\ref{fig:(6)}, the ratio 
$R_{th}(Q^2)\equiv \sigma_L/\sigma_T$ 
is calculated 
using different values for $C_{S_{11}}^L$.
The solid curve denotes the result for $C_{S_{11}}^L=0.65$,
the dashed for $C_{S_{11}}^L=1$, and the dotted for $C_{S_{11}}^L=0.70$.
The maximum ratio is at $Q^2\approx 0.5$ (GeV/c)$^2$, which is 
in agreement with the finding 
of Ref.~\cite{ravndal71}. 

In the following part, we will see that the overestimation of the
longitudinal cross section $\sigma_L$
comes mainly from the photon vertex, where the longitudinal helicity 
amplitude $S_{\frac 12}$ has been overestimated.


\subsubsection{$S_{11}(1535)$ photo- and electro-excitation 
helicity amplitudes}

To better understand the overestimation of the $\sigma_L$, 
we independently calculate $S_{\frac 12}$ of the transition
$\gamma_{(v)} p\to S_{11}(1535)$ in the EVF,
\begin{equation}
\label{longi2}
S_{\frac 12}={\frac 13} \sqrt{\frac{\pi}{\omega_\gamma}} 
\alpha_e^{\frac 12} \frac{|{\bf k}|}{\alpha\gamma_k^3}
e^{-\frac{{\bf k}^2}{6\alpha^2\gamma_k^2}} \ ,
\end{equation}
where $1/\gamma_k$ is the Lorentz boost factor defined in previous Section.

On the other hand, 
we explicitly write out the 
longitudinal cross section $\sigma_L(S_{11})$ for the $S_{11}(1535)$,
\begin{equation}
\sigma_L(S_{11})=\frac{Q^2}{|{\bf k}|^2}\frac{\pi}{\omega_\gamma}
\frac{\alpha_e\alpha_{S_{11}}|{\bf q}| |{\bf k}|^2}
{M_N M_{S_{11}}\Gamma^2_{S_{11}}}\frac{8}{9}
\left[ \frac{\omega_\eta}{\mu_q}-(\frac{\omega_\eta}{E_f+M_N}+1)
\frac{2{\bf q}^2}{3\alpha^2\gamma_q} \right]^2 
\frac{e^{-({\bf k}^2/\gamma_k^2+{\bf q}^2/\gamma_q^2)/3\alpha^2}}
{\gamma_q^4\gamma_k^6} \ ,
\end{equation}
where the Lorentz boost factors have been included 
and the kinematical condition 
$s=M^2_{S_{11}}$ at the $S_{11}(1535)$ c.m. energy has been used.
Substituting the above equation into Eq.~\ref{L},
$S_{\frac 12}$ can be also derived:
\begin{equation}
\label{longi-s11}
|S_{\frac 12}|^2=(C^L_{S_{11}})^2\frac{\pi}{\omega_\gamma}
\frac{\alpha_e\alpha_{S_{11}}|{\bf q}| |{\bf k}|^2}
{M_N^2 b_\eta\Gamma_{S_{11}}}\frac{2}{9}
\left[ \frac{\omega_\eta}{\mu_q}-(\frac{\omega_\eta}{E_f+M_N}+1)
\frac{2{\bf q}^2}{3\alpha^2\gamma_q} \right]^2 
\frac{e^{-({\bf k}^2/\gamma_k^2+{\bf q}^2/\gamma_q^2)/3\alpha^2}}
{\gamma_q^4\gamma_k^6} \ .
\end{equation}

\begin{figure}
\begin{center}
\epsfig{file=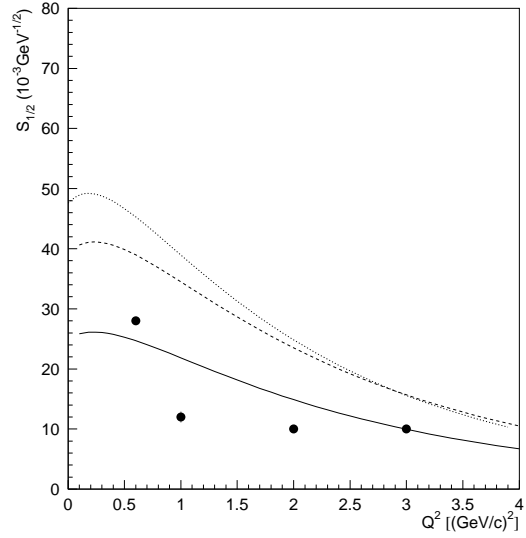,height=8.0cm,width=8.0cm}
\end{center}
\caption{
Longitudinal helicity amplitude $S_{\frac 12}$ for $S_{11}(1535)$.
The solid curve and dashed curves are calculated with $C_{S_{11}}^L=0.65$ 
and $C_{S_{11}}^L=1$, respectively, by Eq.~\ref{L}, while the dotted curve 
is given by Eq.~\ref{longi2}. }
\protect\label{fig:(7)}
\end{figure}

In Fig.~\ref{fig:(7)}, 
the results for $S_{\frac 12}$ from Eq.~\ref{longi2} (dotted curve)
and from Eq.~\ref{longi-s11} 
with $C_{S_{11}}^L=0.65$ (solid curve) and $C_{S_{11}}^L=1$ (dashed curve),
are presented.
Interestingly, it shows that
although the dotted and dashed curves both overestimate
the helicity amplitude $S_{\frac 12}$, 
their values are very close. We thus 
equate these two expressions to give:
\begin{equation}
\label{branching}
b_\eta=\frac{\alpha_{S_{11}}}{\Gamma_{S_{11}}} \chi \ ,
\end{equation}
where
\begin{equation}
\label{chi}
\chi\equiv \frac{2\alpha^2 |{\bf q}|}{M_N^2}
\left[ \frac{\omega_\eta}{\mu_q}-(\frac{\omega_\eta}{E_f+M_N}+1)
\frac{2{\bf q}^2}{3\alpha^2\gamma_q} \right]^2 
\frac{e^{-\frac{{\bf q}^2}{3\alpha^2\gamma_q^2}}}{\gamma_q^4} \ ,
\end{equation}
and 
$\alpha_{S_{11}}\equiv\alpha_\eta ( C_{S_{11}} )^2$
is the $\eta pS_{11}$ couping constant.

Intuitively, the above expression is a reasonable deduction since
Eq.~\ref{branching} is exactly the branching ratio obtained 
by calculating $S_{11}(1535)\to \eta p$. However, recalling that
$b_\eta$ and $\Gamma_{S_{11}}$ are quantities to be investigated
here, relations (as
Eqs.~\ref{branching}-\ref{chi}) 
derived from only $S_{11}(1535)\to \eta p$
cannot help us much, not mention that $\alpha_{S_{11}}$ is also
unclear. We thus expect that information about these quantities
can be derived in
the photo- and electroproduction process.
Unfortunately, another difficulty arises from the electromagnetic 
vertex in photo- and electroproduction. 
Since the experiment cannot separate the $\gamma N S_{11}$
from the strong $\eta N S_{11}$ couplong, studies of $\gamma N\to S_{11}$
and $S_{11}\to \eta N$ are strongly model-dependent~\cite{li-close-90}.
Therefore, the fit of the photo- and electroproduction data 
generally tells us what is the ``parameter" (e.g. $\alpha_{S_{11}}$) 
for $\gamma N\to S_{11} \to \eta N$ instead of  
what is the coupling for $S_{11}\to \eta N$.
The overestimation of the $S_{\frac 12}$ is an example
exposing this kind of problems. 

Now, the trivial-look Eqs.~\ref{branching} and \ref{chi}
can tell us more about the $\eta N S_{11}$ coupling.
First, since these quantities, $b_\eta$, $\Gamma_{S_{11}}$,
and $\alpha_{S_{11}}$, are determined by fitting the reaction,
we thus should bear in mind that their values have
contained uncertainties from the electromagnetic interactions.
Also, the model-dependent feature
arising from non-$S_{11}$ contributions could mix into 
the fitted values for those parameters.
As shown by Fig.~\ref{fig:(7)}, 
although the fitted quantity $C^L_{S_{11}}=0.65$
better accounts for the data, we obviously cannot equate the
solid curve to the dotted one. 
In that case, Equations~\ref{branching} and \ref{chi}
would not be a natural deduction at all.
The non-trivial thing occurs only if we take all parameters fitted by the data
except for $C^L_{S_{11}}=1$ to produce the dashed curve in Fig.~\ref{fig:(7)}. 
Namely, both overestimations of the 
electromagnetic coupling turn to be equal to each other.
Thus, it leads to the relation of Eq.~\ref{branching}, in which 
the electromagnetic coupling has been cleanly removed.
Since on the one hand, 
no information about the electromagnetic interaction
is contained in Eqs.~\ref{branching} and \ref{chi} any more, and 
on the other hand, all quantities involved are determined
by the fits, these two equations in effect provide an additional 
relation for all those fitted quantities.
We thus can ask 
the question concerning the self-consistence 
of the model fitting results, {\it whether
those quantities fitted by the reaction data can satisfy 
the relation}? Again, we have to bear in mind that those 
quantities could have contained all uncertainties 
arising from the electromagnetic coupling as well as 
model-dependent features from the non-$S_{11}$ terms.

To answer the question, we calculate the
partial decay width of the $S_{11}(1535)$ using those fitted quantities, 
i.e. $\Gamma^a_\eta\equiv b_\eta\Gamma_{S_{11}}$, and 
$\Gamma^b_\eta\equiv \alpha_{S_{11}} \chi$, where 
$\chi$ is calculated by Eq.~\ref{chi}. 
The results are presented in Table~\ref{tab:(2)}. 
The PDG~\cite{PDG98} estimations are also presented. 
It shows that $\Gamma^a_\eta$ and $\Gamma^b_\eta$
are in good agreement with each other, although
discrepancies also exist. 
Quite clearly, the discrepancies reveal the
uncertainties arising from the model-dependent 
aspects in the fits, while the consistencies 
highlight a reasonable constraint to the strong coupling.
It suggests that the effective Lagrangian successfully accounts 
for the $\eta N S_{11}$ coupling within an accuracy of 15\%. 
We shall see below that this conclusion is important
for the assessment  of the $S_{11}$ transverse helicity amplitude 
derived in the reaction. 
Note that the harmonic oscillator 
strength $\alpha$ has a moderate value 384.5 MeV,
which is within the range of commonly used values, 330-430 MeV,
this provides another justification
for this approach. In this sense, 
the relation of Eq.~\ref{branching} 
could be regarded as being ``satisfied" 
rather than ``fitted".

\begin{figure}
\begin{center}
\epsfig{file=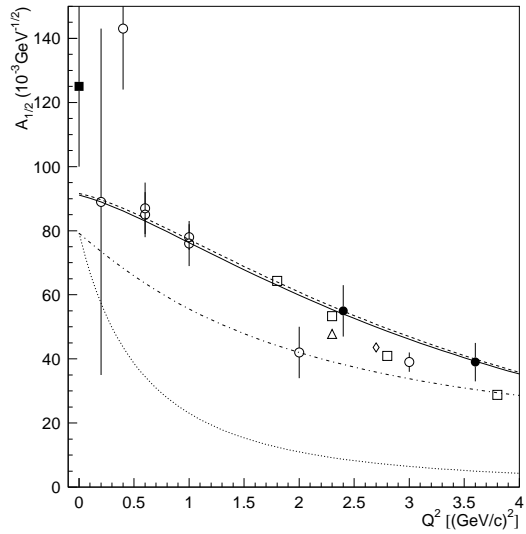,height=8.0cm,width=8.0cm}
\end{center}
\caption{
Transverse helicity amplitude $A_{\frac 12}$ 
for the $S_{11}(1535)$. The solid and dashed curves 
are calculated by Eqs.~\ref{heli-2} and ~\ref{heli-1}, respectively.
The dot-dashed curve is calculated based on 
the Lorentz boost factor in the meson-nucleon c.m. frame, and 
the dotted curve based on a dipole form factor.
The full circles are from 
Ref.~\protect\cite{armstrong99} and other data points from 
Refs.~\protect\cite{mainz,stoler93,Br-76}. }
\protect\label{fig:(8)}
\end{figure}

In the following part, we turn to the results for the transverse 
helicity amplitude for the $S_{11}(1535)$. 
From Eq.~\ref{T}, the transverse helicity amplitude for the $S_{11}(1535)$
can be expressed as 
\begin{eqnarray}
\label{heli-1}
|A_{\frac 12}|^2 &= &\frac{M_{S_{11}}}{2 M_N}\frac{\Gamma_{S_{11}}}{b_\eta}
\sigma_T(\gamma_{(v)} p \to S_{11} \to \eta p) \nonumber\\
&=&\frac{\pi}{\omega_\gamma}
\frac{\alpha_e\alpha_{S_{11}}|{\bf q}|}
{M_N^2 b_\eta\Gamma_{S_{11}}}
\frac{(E_f+M_N)}{2M_N}\nonumber\\
&&\times\left(\omega_\gamma+\frac{{\bf k}^2}{2m_q\gamma_k}\right)^2
\frac{e^{-({\bf k}^2/\gamma_k^2+{\bf q}^2/\gamma_q^2)/3\alpha^2}}
{\gamma_q^4\gamma_k^6}\nonumber\\
&&\times\frac{2}{9}\left[ \frac{\omega_\eta}{\mu_q}
-\left(\frac{\omega_\eta}{E_f+M_N}+1 \right)
\frac{2{\bf q}^2}{3\alpha^2\gamma_q} \right]^2 \ ,
\end{eqnarray}
where $\sigma_T(\gamma_{(v)} p \to S_{11} \to \eta p)$ is the exclusive 
transverse cross section calculated by the model, and the Lorentz
boost factors are taken into account. 
The result of Eq.~\ref{heli-1} is presented 
in Fig.~\ref{fig:(8)} (dashed curve),
which is in good agreement with the experimental data.
It is worth noting that 
a direct relation as Eq.~\ref{branching} cannot be obtained 
for the transverse helicity amplitude. The basic reason is that
the EM operators used in this model~\cite{li97} are essentially different 
from the NRCQM ones~\cite{isgur-karl}. 
However, as shown 
by the dashed curve in Fig.~\ref{fig:(8)}, 
those well constrained parameters indeed provide a reasonable 
discription of the $S_{11}(1535)$. 

Such a result certainly highlights again the relation 
provided by Eq.~\ref{branching} and \ref{chi}.
It is also useful for the understanding of a variable 
defined in the literature. 
In Ref.~\cite{RPI96}, it was found that the quantity
$\xi_T\equiv\sqrt{\Omega\Gamma_\eta} A_{\frac 12}/\Gamma_{S_{11}}$, 
where $\Omega$ is the phase space factor, was not sensitive 
to the fitting scheme since uncertainties within $\Gamma_\eta$
and $\Gamma_{S{11}}$ went to the same direction.
That feature arises because the constraint to the $\eta N S_{11}$ vertex
was not enough. However, here we find that  
Eq.~\ref{branching} restricts the $b_\eta$ and $\Gamma_{S_{11}}$
in an inverse direction compared to $\xi_T$. Supposing 
$b_\eta$ and $\Gamma_{S_{11}}$ both increase  or decrease, 
the relation of Eq.~\ref{branching} will be destructively unbalanced. 
The satisfactory of Eq.~\ref{branching} thus indeed
serves as a further constraint to those fitted quantities.

The dominant $S_{11}(1535)$ production can be investigated by 
substituting the exclusive $\sigma_T(\gamma_{(v)} p \to S_{11} \to \eta p)$ 
with the inclusive $\sigma_T$ in Eq.~\ref{heli-1}:
\begin{equation}
\label{heli-2}
|A_{\frac 12}|^2=\frac{M_{S_{11}}}{2 M_N}\frac{\Gamma_{S_{11}}}{b_\eta}
\sigma_T \ .
\end{equation}
The difference between Eqs.~\ref{heli-1} and ~\ref{heli-2} reflects 
the influence of the non-$S$-wave contributions in the transverse helicity
amplitude. 
The calculations of Eq.~\ref{heli-1} (dashed curve), 
and Eq.~\ref{heli-2} (solid curve) are presented in Fig.~\ref{fig:(8)}.
It shows that the impact 
of the non-$S_{11}$ contributions produces about 3\% difference
in the helicity amplitude, and the transverse cross sections 
are dominated by the $S_{11}(1535)$ excitation.

In Ref.~\cite{armstrong99}
the impact from the longitudinal cross section have been investigated through 
the following relation:
\begin{equation}
\label{heli-3}
|A_{\frac 12}|^2=\frac{M_{S_{11}}}{2 M_N}\frac{\Gamma_{S_{11}}}{b_\eta}
\frac{\sigma_T(\gamma p \to S_{11} \to \eta p)}{(1+\varepsilon R(Q^2))} \ .
\end{equation}

As shown in Fig.~\ref{fig:(5)}, 
the ratio $R(Q^2)$ is much smaller than unit, thus, 
the effect of the 
longitudinal contribution, reflected by the factor $1/(1+\varepsilon R(Q^2))$, 
is found negligible in comparison with
Eq.~\ref{heli-1}, especially at high $Q^2$ regions.

In Fig.~\ref{fig:(8)}, we also present the calculation of $A_{\frac 12}$
based on the dipole-like Lorentz boost factor in the 
meson-nucleon c.m. frame (Eq.~\ref{breit-photon}) 
for the $S_{11}(1535)$ (See the dot-dashed curve). 
With the same set of parameters, the dot-dashed curve 
apparently under-estimates the experimental values, although 
compared to a parametrized dipole form factor,
$1/(1+Q^2/0.8)^2$ (see the dotted curve),
it better accounts for the trend of data.

\begin{figure}
\begin{center}
\epsfig{file=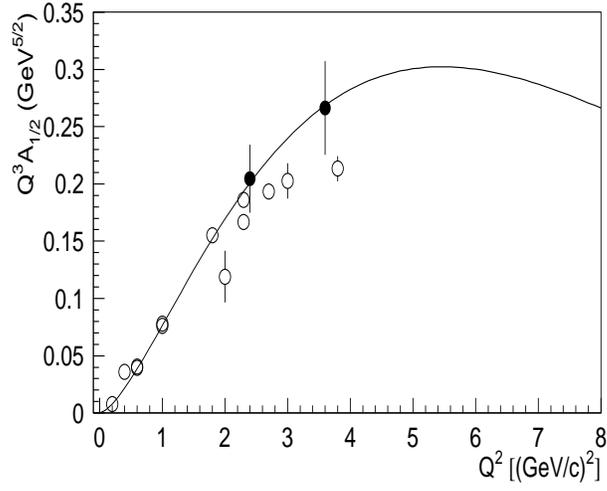,height=8.0cm,width=10.0cm}
\end{center}
\caption{ 
The quantity $Q^3A_{\frac 12}(Q^2)$ calculated in this model. 
The full circles are from 
Ref.~\protect\cite{armstrong99} and the empty ones from 
Refs.~\protect\cite{mainz,stoler93,Br-76}. }
\protect\label{fig:(9)}
\end{figure}

The resonance contributions to the transverse helicity amplitude 
$A_{\frac 12}(Q^2)$ drop down with the increasing $Q^2$ 
as shown by the full curve in Fig.~\ref{fig:(8)}. 
At high $Q^2$ region, the pQCD dominance in the
helicity-conserved amplitude $A_{\frac 12}(Q^2)$ predicted
the asymptotic behavior of $A_{\frac 12}(Q^2)$ fall-off 
like $1/Q^3$~\cite{carlson88}. 
That is to say, the quantity $Q^3A_{\frac 12}(Q^2)$ 
will approach a constant at high $Q^2$ regions. 
However, the studies by Isgur and Smith~\cite{isgur84}, 
and Radyushkin~\cite{radyushkin91} led to quite different results. 
They found that such a scaling behavior in the exclusive processes 
were still dominated by the non-perturbative contributions rather 
than the perturbative ones over wide $Q^2$ regions.
We present the calculation of the quantity $Q^3A_{\frac 12}(Q^2)$ 
in Fig.~\ref{fig:(9)} to compare with the data. 
At $Q^2=2.4$ and 3.6 (GeV/c)$^2$, 
the quantity $Q^3A_{\frac 12}(Q^2)$ is calculated through Eq.~\ref{heli-1}. 
It gives 0.201 and 0.268 GeV$^{\frac 52}$, respectively, which are 
consistent with the non-scaling behavior found in experiment at small $Q^2$. 
At $Q^2=4.0$ (GeV/c)$^2$, we get $Q^3A_{\frac 12}=0.282$ GeV$^{\frac 52}$.
This quantity approaches the maximum 
0.302 GeV$^{\frac 52}$
at $Q^2=5.5$ (GeV/c)$^2$, and then falls down slowly with the increasing $Q^2$. 
However, due to the shortcoming of the NRCQM, the results 
above 4 (GeV/C)$^2$ should not be taken seriously. 
More fundamental approach is needed for the dying-out region of 
non-purterbative processes, where a continuing 
scaling behavior from pQCD processes might emerge.


\section{Summary and conclusion}

This work was motivated by the recent electroproduction data from 
JLab~\cite{armstrong99}
in the $S_{11}$ dominant kinematical region. We also took advantage
of even more recent photoproduction data from GRAAL~\cite{graal-coll}. 
Such a database,
expected to be significantly enlarged in the near future, offers an 
excellent opportunity to investigate the reaction mechanism of the $\eta$
meson photo- and electroproduction. 
In particular, the selected kinematical region allows us to study the 
nature of the $S_{11}(1535)$ resonance with unprecedently precise 
data. 

The model presented here is an extension of a constituent 
quark model approach with a chiral effective Lagrangian
~\cite{li97} to the 
electroproduction process. 
On the basis of the symmetric NRCQM,
there are only a limited number of parameters appearing in this
approach. 
In principle, the breaking of this symmetry can be phenomenologically
introduced through an additional coefficient for each resonance, 
which then could be determined in the numerical 
study~\cite{li-saghai98,saghai-li-01}. 
However, different from Refs.~\cite{li-saghai98,saghai-li-01},
in this work attentions are paid to the $S_{11}(1535)$ and $D_{13}(1520)$.
The selected kinematics permit us to introduce the symmetry breaking 
coefficients only for the $S_{11}(1535)$ and $D_{13}(1520)$ due to 
the dominance of the former and significant interferences from the latter. 
For other resonances, their symmetry breaking 
coefficients are neglected due to their negligible effects.
In this work, the $S_{11}(1535)$'s total width, branching ratio of 
$\eta N$ channel, as well as the harmonic oscillator strength 
of the quark model are treated as parameters and determined by the fits.
The quark model symmetry breaking could introduce
contributions from the resonance $S_{11}(1650)$ and $D_{13}(1600)$, 
which have been suppressed in the SU(6)$\otimes$O(3) symmetry limit
due to the Moorhouse selection rule~\cite{moorhouse}.
Although an explicit consideration of such a breaking 
has not been doen here, a possible influence 
of these resonances could have been 
contained in those fitted quantities as a background contribution.
In this sense, the 15\% uncertainty for the $\eta N S_{11}(1535)$
coupling can be regarded as a reasonable estimation.
Another feature of this approach is that 
the Lorentz boost factor in the equal velocity frame 
(EVF) succeeds in the description of the $S_{11}(1535)$ form factor.
Therefore, an {\it ad hoc} form factor is avoided.

We have shown that the extracted parameters are compatible with
their values coming from previous independent works. 
The cross section data for both photo- and electroproduction 
are well reproduced. 
Explicitly, it shows that the electroproduction process
indeed provides us with rich information. It sheds light on the 
$\eta N S_{11}(1535)$ coupling, which had not been
well-determined in the real photon reaction. 
Although the scalar coupling 
of the longitudinal photon is found to be larger than that reported 
in Ref.~\cite{breuker78}, we succeed in removing 
the model-dependent uncertainties 
arising at the electromagnetic 
interaction vertex ($\gamma_{(v)} p \to S_{11}$) from 
the transition amplitude $\gamma_{(v)} p \to S_{11} \to \eta p$
for the $S_{11}(1535)$.
The derived simple relation (Eq.~\ref{branching}) thus provides 
a constraint on the $\eta N S_{11}$ coupling with 
an uncertainty of about 15\%.  
This could be the first explicitly constrained estimation
of the $S_{11}(1535)$ properties in theory.
Such a result might lead us to the conclusion that 
the main component in the $S_{11}(1535)$ wave function
is rather like a non-exotic three-quark state than a  
$K\Lambda$ ($K\Sigma$) bound state~\cite{zr96}. 
Possible configurations
from the latter  
should be small at the $S_{11}(1535)$ energy, if exists. 

It should be noted that this calculation was done before 
the publication of a new set of data from the CLAS
Collaboration~\cite{clas2001}. 
The $S_{11}$ helicity 
amplitude $A_{\frac 12}$ for 0.25 $\le Q^2 \le$ 1.5 (GeV/c)$^2$
has been significantly improved by that experiment. 
Their results, which 
are highly model-selective, shows in good agreement 
with our calculations.

\acknowledgments
We are grateful to the GRAAL collaboration, especially J.-P. Bocquet, 
D. Rebreyend, and F. Renard for having provided us with their data 
prior to publication. We wish to thank J.-P. Didelez, M. Guidal and
E. Hourany for useful discussions concerning the GRAAL experiments and
S. Dytman, B. Ritchie and P. Stoler for
helpful exchanges on the measurements performed at JLab.
This work is supported by the 
``Bourses de Recherche CNRS-K.C. WONG'' and IPN-Orsay.

%

%
%
%
\begin{table}
\caption{ 
Extracted values for parameters of this model. 
Detailed discussions are given in the text. }
\protect\label{tab:(1)}
\begin{center}
\begin{tabular}{cccccccc}
$\alpha_\eta $  & $\alpha$ (MeV) & 
$\Gamma_{S_{11}}$ (MeV)& $b_\eta(S_{11})$ 
& $C_{S_{11}}$   & $C_{D_{13}}$    & $C_{S_{11}}^L$ & $\chi^2$ \\[1ex]
\hline\\[1ex]
$0.10\pm 0.01$ & $384.5\pm 0.2$ & $143.3\pm 0.2$    & $0.55\pm 0.01$   
& $1.20\pm 0.30$ & $-0.92\pm 0.03$ & $0.65\pm 0.01$ & $2.36$ \\[1ex]
\end{tabular}
\end{center}
\end{table} 
 
%
%
%
\begin{table}
\caption{ Extracted total width $\Gamma_{S_{11}}$ and
partial decay width $\Gamma_\eta^a\equiv b_\eta \Gamma_{S_{11}}$ 
and $\Gamma_\eta^b\equiv \alpha_{S_{11}}\chi$ for the $S_{11}(1535)$.
Corresponding estimations by the 
Particle Data Group~\protect\cite{PDG98} are also listed. }
\protect\label{tab:(2)}
\begin{center}
\begin{tabular}{cccc}
 & $\Gamma_{S_{11}}$ (MeV) & $\Gamma_\eta^a$ (MeV) & 
 $\Gamma_\eta^b$ (MeV)\\[1ex]\hline 
\\[1ex]
 This work &  $143.3\pm 0.2$ &  78.8 & 89.4 \\[1ex]
 PDG       &  150  &  75.0 & - \\[1ex]
\end{tabular}
\end{center}
\end{table}    
%
%
%
%
%
%
%
%
%
%
%
%
%
%
%
%
%
%
%

\end{document}